\documentclass[sigconf]{acmart}

\usepackage[linesnumbered,lined,boxed,commentsnumbered]{algorithm2e}
\usepackage{subcaption}
\usepackage[nolist,nohyperlinks]{acronym}
\usepackage{todonotes}

\acrodef{SDI}[SDI]{\emph{Spatial Distancing Index}}
\acrodef{CFD}[CFD]{\emph{Computational Fluid Dynamics}}

\DeclareMathOperator*{\argmin}{arg\,min}

\AtBeginDocument{%
  \providecommand\BibTeX{{%
    \normalfont B\kern-0.5em{\scshape i\kern-0.25em b}\kern-0.8em\TeX}}}

\setcopyright{acmcopyright}
\copyrightyear{2021}
\acmYear{2021}
\acmDOI{}

\acmConference[SimAUD '21]{SimAUD '21: Proceedings of the 12th Annual Symposium on Simulation for Architecture and Urban Design}{April 15-17, 2021}{Virtual Event, USC}
\acmBooktitle{SimAUD '21: Proceedings of the 12th Annual Symposium on Simulation for Architecture and Urban Design,
   April 15-17, 2021, Virtual Event, USC}
\acmPrice{}
\acmISBN{}

\begin{document}

\title{Optimizing Indoor Navigation Policies For Spatial Distancing}

\author{Xun Zhang}
\email{xz348@scarletmail.rutgers.edu}
\affiliation{%
  \institution{Rutgers University}
  \city{New Brunswick}
  \state{New Jersey}
  \country{USA}
}

\author{Mathew Schwartz}
\email{cadop@njit.edu}
\orcid{0000-0003-3662-7203}
\affiliation{%
  \institution{New Jersey Institute of Technology}
  \city{Newark}
  \state{New Jersey}
  \country{USA}
}

\author{Muhammad Usman}
\email{usman@cse.yorku.ca}
\affiliation{%
  \institution{York University}
  \city{Toronto}
  \state{Ontario}
  \country{Canada}
}

\author{Petros Faloutsos}
\email{pfal@eecs.yorku.ca}
\affiliation{%
 \institution{York University\\
 UHN Toronto Rehabilitation Institute}
 \city{Toronto}
 \state{Ontario}
 \country{Canada}
 }

\author{Mubbasir Kapadia}
\email{mk1353@cs.rutgers.edu}
\affiliation{%
  \institution{Rutgers University}
  \city{New Brunswick}
  \state{New Jersey}
  \country{USA}
}

\renewcommand{\shortauthors}{Xun, et al.}


\begin{abstract}

The ability for occupants to shop, relax, or work is facilitated by not only the objects in the space but the arrangement and movement patterns allowed. In this regard, we consider one type of affordance in a space; the ability to maintain distance between occupants while navigating and interacting with the environment. In this paper, we focus on the modification of policies that can lead to movement patterns and directional guidance of occupants, which are represented as agents in a 3D simulation engine. Considering an environment can dictate –- or even mandate -- the directional movement of the occupants, we demonstrate an optimization method that improves a spatial distancing metric by modifying the navigation graph. To achieve this, we introduce a measure of spatial distancing of agents as a function of agent density (i.e., occupancy). Our optimization framework utilizes such metrics as the target function, using a hybrid approach of combining genetic algorithm and simulated annealing. We show that within our framework, the simulation-optimization process can help to improve spatial distancing between agents by optimizing the navigation policies for a given indoor environment.

\end{abstract}

\ccsdesc[300]{Human-centered computing~Accessibility design and evaluation methods}
\ccsdesc[500]{Applied computing~Computer-aided design}
\ccsdesc[100]{General and reference~General conference proceedings}

\keywords{Graph optimization, genetic algorithm, simulated annealing, spatial distancing index, navigation, multi-agent simulation}

\begin{teaserfigure}
  \includegraphics[width=\textwidth]{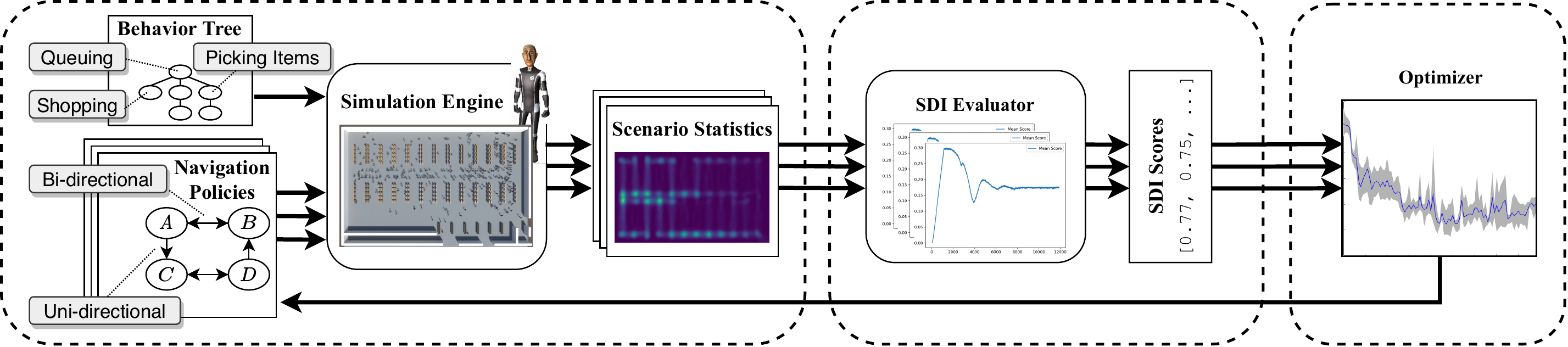}
  \caption{Framework overview. Our framework consists of three main components: (i) crowd navigation and behavior simulation, (i) a model for spatial distancing index, and (iii) a graph optimizer.}
  \Description{Framework overview. Our framework consists of three main components: (i) crowd navigation and behavior simulation, (i) a model for spatial distancing index, and (iii) a graph optimizer.}
  \label{fig:overview}
\end{teaserfigure}

\maketitle

\section{Introduction} \label{sec:intro}
While many features of a building are quantifiable through existing simulation protocols, crowd management is an ongoing and difficult aspect of occupant experience to analyze. Although past work has studied simulation-based locomotion of agents in an environment, there is a challenge in dictating movement policies that facilitate the best outcome, such as maintaining distance, for a given space.  

With the increased recognition of human health and safety, there is an increasingly important role of simulating the experience and conditions of humans themselves in the built environment, compared to the natural conditions such as lighting and heating.  These simulations can aid during the design process, but importantly, in the post-occupancy analysis as well. Unlike typical human subject-based post-occupancy evaluations, questions such as: \textit{how can we minimize contact of people?} are not feasibly done with people, and instead, we can leverage simulation to aid in the decision process.  

While existing work has focused heavily on either the control of crowd simulation or the use of crowd simulation for human-centered performance criteria, few works have considered this system from a management perspective in which generic navigation guidelines or policies are applied to the crowd.

In this paper, we present a novel graph-based model that describes the crowd management policies of an environment. Then, we conduct crowd simulations driven by these policies and collect temporal and spatial statistics of the agents' locomotion in the environment. Next, we evaluate the environment through a \ac{SDI} value based on the crowd simulation statistics. Finally, we formulate the graph optimization problem using the \ac{SDI} measuring and devise a novel optimization algorithm to find the best applicable policy of this environment. We demonstrate the efficacy of our method in several case studies to showcase the validity of our measure for social distancing, and the potential of using our framework to discover optimal navigation policies.

\section{Related Work} \label{sec:related}

The growing research area combining crowd modeling and human behavior simulation has led to various examples of how the complexities of occupant movement and interaction with the built environment can provide valuable feedback on design choices.

Simulating crowds has been a topic of interest for many years in the game and animation industry~\cite{reynolds1999steering}
. Two key aspects of crowd simulation that must be considered, especially for building evaluation, are the agents' realistic behavior and decision-making abilities in the simulation. Numerous studies have considered methods for implementing human behaviors such as wayfinding tasks \cite{dubey2020autosign}, narratives \cite{schaumann2017simulating}, panic \cite{helbing2002crowd}, building evacuations~\cite{helbing2013pedestrian}, among others. Likewise, the validity of crowd behaviors has been studied~\cite{banerjee2010evaluation}
, showing varying levels of accuracy to real-world trials. While there is no single simulator that perfectly matches all conditions, the estimation of crowd behavior in simulation is well accepted for important decisions such as egress \cite{cassol2017evaluating}, wayfinding \cite{huang2017automatic}, and to analyze building environments for human occupancies in general~\cite{usman2018interactive}. It has also been used with numerous commercial software products in existence (e.g., mass motion~\cite{massmotion}, and anylogic~\cite{anylogic}).

Various crowd-based measures can be extracted from the simulations related to the occupant experience and quality of the building design. For example, crowd flow measures the rate at which the agents pass through a given point of interest in the environment (e.g., doors), crowd trajectories yield the path that individual agents follow during the course of the simulation, and evacuation time measures the egress time of the agents. These measures have been widely used to analyze crowd dynamics in building designs~\cite{berseth2015environment}.

Recently, researchers have been using optimization techniques to optimize various contexts in building design, from lighting layouts~\cite{nagy2017project}, furniture~\cite{weiss2019position,yu2011make}, building layouts \cite{feng2016crowd,das2016space}, to urban networks \cite{nagy2018generative,miao2020development, chirkin2016concept}. With the global impact of COVID-19~\cite{WHOsituation} forcing new policies of maintaining distance~\cite{WHOSocialDistancing} and the economic impacts of complete lockdowns~\cite{coibion2020cost}, it has become of great interest the ability to balance the number of people in a space with maximal distance. 
Recent works have used crowd simulation to determine probabilities of infection and spread~\cite{unitysim}, and to evaluate the impact of an environment's layout for violations to a desired occupant distancing rule~\cite{usman2020social}. However, the idea of the distance between occupants is not isolated to disease. More generally, psychological, social, and cultural factors influence the comfort levels of people based on their distances~\cite{hall1968proxemics}.

The work presented in~\cite{pettre2008crowds} used geometry analysis to deduce the environment structure (e.g., topological and geometry information) as a navigation graph, which is then used as the basis for navigation planning and simulation. Graph optimization using simulated annealing has been studied thoroughly \cite{johnson1989optimization}
, while applications regarding crowd control or policy optimization have rarely been conducted. Simulated annealing has been used for the graph and network optimization in various fields such as traffic \cite{zhao2006simulated}. Engineering control for indoor environments has also been studied to control the transmission of airborne diseases, while the studied factors are limited, mainly related to airflow control and ventilation \cite{MORAWSKA2020105832}. 
Recently,~\cite{unitysim} demonstrated a method to simulate various combinations of a grocery store configuration using navigation graphs and shopping lists. This was to reduce disease spread based on their disease transmission model, while their optimization is a brute-force approach.

The use of an index for evaluating environments (built or virtual) has been of interest and use in the past. There is a walkability index \cite{frank2010development} that is commonly used for urban planning and Indoor Walkability Index \cite{shin2019indoor} for building circulation evaluation. We continue this along this human-centric theme by devising an index scoring, or ranking, a buiding's design through the \ac{SDI}. Further details on the formulation and calculation of \ac{SDI} can be found in Section~\ref{sec:sdi}.

\section{Overview and Preliminaries} \label{sec:overview}
Our framework consists of 3 major components: (i) the crowd behavior simulation engine, (ii) the SDI model, and (iii) the graph optimizer. An overview of the complete working pipeline is shown in Figure \ref{fig:overview}. A list of terms used is presented in Table~\ref{tab:term}.

A complete cycle starts with an environment guidance policy, described by a \textit{structural graph}. Along with the necessary behavior trees, the structural graph is then utilized by the simulation engine and a corresponding \textit{navigational graph} is generated. Throughout the simulation, statistical data are collected and sent to the SDI evaluator to calculate the score of the current policy. As the last step in the cycle, the score is sent to the optimizer, which decides the next optimization step to take. 

\begin{figure*}[!htbp]
  \centering
  \includegraphics[width=.95\linewidth]{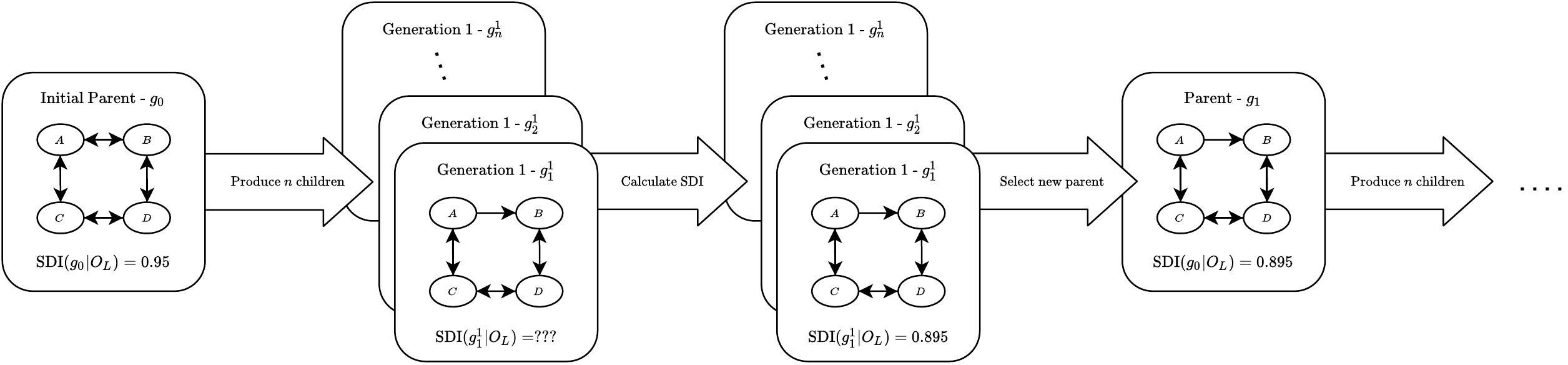}
  \caption{The simulation-optimization pipeline. For each graph in a generation, we run a simulation session and \ac{SDI} is calculated. We then use the graph policy with the minimum \ac{SDI} value as the parent graph for the next generation or reject it with a decreasing probability. The optimization process terminates if the \ac{SDI} value meets the set convergence threshold.}
  \label{fig:optim_pipe}
\end{figure*}

\subsection{Graph-based Representation of Crowd Management Policies}
\label{subsec:graph-based-representation}

We propose a graph representation of the directional policies enforced in the environment using a tuple of directed graphs $<g,h>$. The graphs are represented with nodes and directed edges. Each tuple of graphs contains (i) a \textit{structural} graph $g$ and (ii) a \textit{navigational} graph $h$.

\begin{table}
    \centering
    \small
    \begin{tabular}{c|c||c|c}
        \hline
        Symbol & Notation & Symbol & Notation\\
        \hline
        $h$ & Time factor & $m$ & Min distance\\ 
        $q$ & Max distance & $t$ & Time\\
        $O_{L}$ & Occupancy load & $e, e_{ij}$ & Edge\\
        \hline
        $g,g_c$ & Structural graph & $G,G_c$ & Set of $g,g_c$\\ 
        $h$ & Navigational graph & $H$ & Set of $h$\\
        $n,n_i$ & Node & $N,N_g,N_h$ & Set of $n,n_i$\\
        \hline
    \end{tabular}
    \caption{Symbols used in this paper.}
    \label{tab:term}
\end{table}

For each environment, a \textit{structural} graph is a static and abstract model that describes the high-level features of the environment, i.e., the directional policies enforced on the edges and the key nodes in the environment. Such key nodes in the structural graph represent significant locations in the environment, e.g., the entrance, the exit, or a joint location of paths. An edge in the graph is considered as an abstract relation between two nodes, i.e., a policy that allows the agent to pass from one node to the other, or forbids such motion. These policies enforce that traveling between any two nodes can only be either bi-directional, single-way, or blocked (disjoint).

\begin{figure}[htbp]
  \begin{subfigure}[b]{\linewidth}
    \centering
    \includegraphics[width=0.8\linewidth]{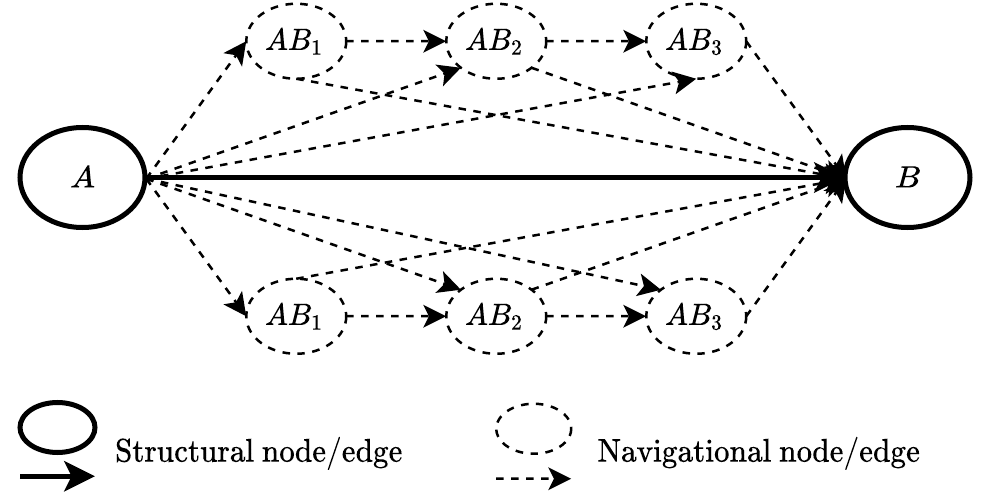}
    \label{subfig:navgraph}
  \end{subfigure}
  \par \medskip
  \begin{subfigure}[b]{\linewidth}
    \centering
    \includegraphics[width=0.75\linewidth]{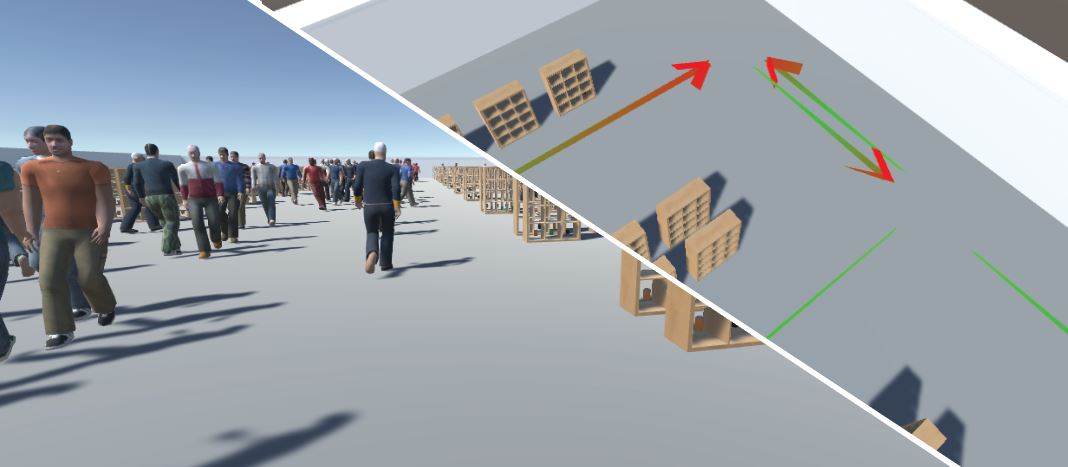}
    \label{subfig:splitscreenagent}
  \end{subfigure}
  \caption{(Top) Example of nodes and edges for the structural and navigational graphs. There are $2$ structural nodes $A$ and $B$ with a directional edge from $A$ to $B$. Along this edge, there are $6$ navigational nodes ($AB_1,AB_2\ldots,AB_6$) and $16$ corresponding navigational edges connecting them. (Bottom) During simulation, the agents will follow the direction for pathfinding. The colored arrows represent the directional policies (e.g., $\textit{Green} \rightarrow \textit{Red}$).}
  \label{fig:graphs}
\end{figure}

Based on the structural graph, we dynamically generate a corresponding \textit{navigational} graph during runtime for the pathfinding of the agents. 
Nodes in the navigational graph contain detailed navigation points of interest such as a specific shopping item. The navigational edges enumerate all possible paths between any two nodes in the structural graph so that when any agent traverses from one node to another, they can also approach a target shopping item without violating the policies. Figure \ref{fig:graphs} shows the relationship and difference between the structural and navigational graphs.

\subsection{Human Behavior Simulation} \label{subsec:hbs}

The human behavior simulation (also known as crowd simulation) uses a Unity-based framework, ADAPT~\cite{kapadia2014adapt}, to design and author meaningful animation, navigation, and agent behaviors in virtual environments. The framework consists of three layers: (i) a goal-directed locomotion and collision avoidance layer, (ii) a navigation layer for path planning, and (iii) a behavior tree layer for parameterized agent behaviors.

\subsubsection{Locomotion and collision avoidance.} This layer controls the steering capabilities of the agents in the virtual environment. It uses ORCA (Optimal Reciprocal Collision Avoidance)~\cite{van2011reciprocal} to find the optimal velocities for the agents using gradient descent. It communicates with the navigation layer for agents' path planning. Certain crowd steering configurations such as walking speed, stopping conditions, and maintaining social distances are also set in this layer.

\subsubsection{Path planning.} The planning layer uses Unity's AI system for pathfinding in the environment. To compute the path from any two nodes in the graph, our nagivational graph first breaks the path down into small trajectory snippets, then for each trajectory snippet, we provide its starting and ending position to Unity's navigation system to conduct the actual locomotion. The navigation graphs are meticulously designed that any trajectory snippet is small enough to prevent the agents from violating the policies.

\subsubsection{Behavior tree.} The behavior tree layer defines all the permissible actions that agents can perform during the simulation. These actions include walking, shopping, queuing, and entering/exiting the environment. The behavior tree layer is also responsible for scheduling and managing the sequence of behaviors that agents carry out during the simulation in an ordered fashion. Please refer to~\cite{shoulson2011parameterizing} for more details on the underlying implementation and authoring of behavior trees. \newline

In the simulation process, we first load the environment model. All the furniture items (e.g., item shelves and checkout counters) are then procedurally generated. Next, the agents are initialized at the entrances. Initially, there is no agent present in the environment. We gradually let the agents enter the facility at a varying rate until the capped maximum capacity of the environment is reached. Afterward, a new agent is spawned only when an existing agent leaves the environment after performing the assigned behaviors. When an agent is spawned in the environment, it is given a list of randomly generated shopping items. The agents' tasks are to shop all the items from the list by traversing the environment and obeying any enforced navigation policies. Once all the items are obtained, the agent queues up at the checkout counter to pay for the items, and exits the environment. In our experiments, we scaled up the simulation speed for higher runtime efficiency.

\subsection{Spatial Distancing Index} \label{subsec:sdi}

We present \ac{SDI} as a function of the occupancy load over time for the given environment. The \ac{SDI} considers the ability to maintain distances between the occupants, and for a given design (and its navigation policy), the ability to maintain distances at the various occupancy loads allowed in the environment space. For more details, see Section~\ref{sec:sdi}.

\subsection{Optimization}

Based on the graph representation of the policies, we devise a hybrid optimization process using the genetic algorithm (GA) and simulated annealing (SA) to approach the optimal policy with the best value of SDI or another arbitrary measuring metric. Details about the optimization framework and implementation of the algorithms are elaborated in Section \ref{optimization}.

\section{Spatial Distancing Index} \label{sec:sdi}

\begin{figure*}[t]
  \centering
  \begin{subfigure}[t]{.33\linewidth}
    \centering
    \includegraphics[width=\linewidth]{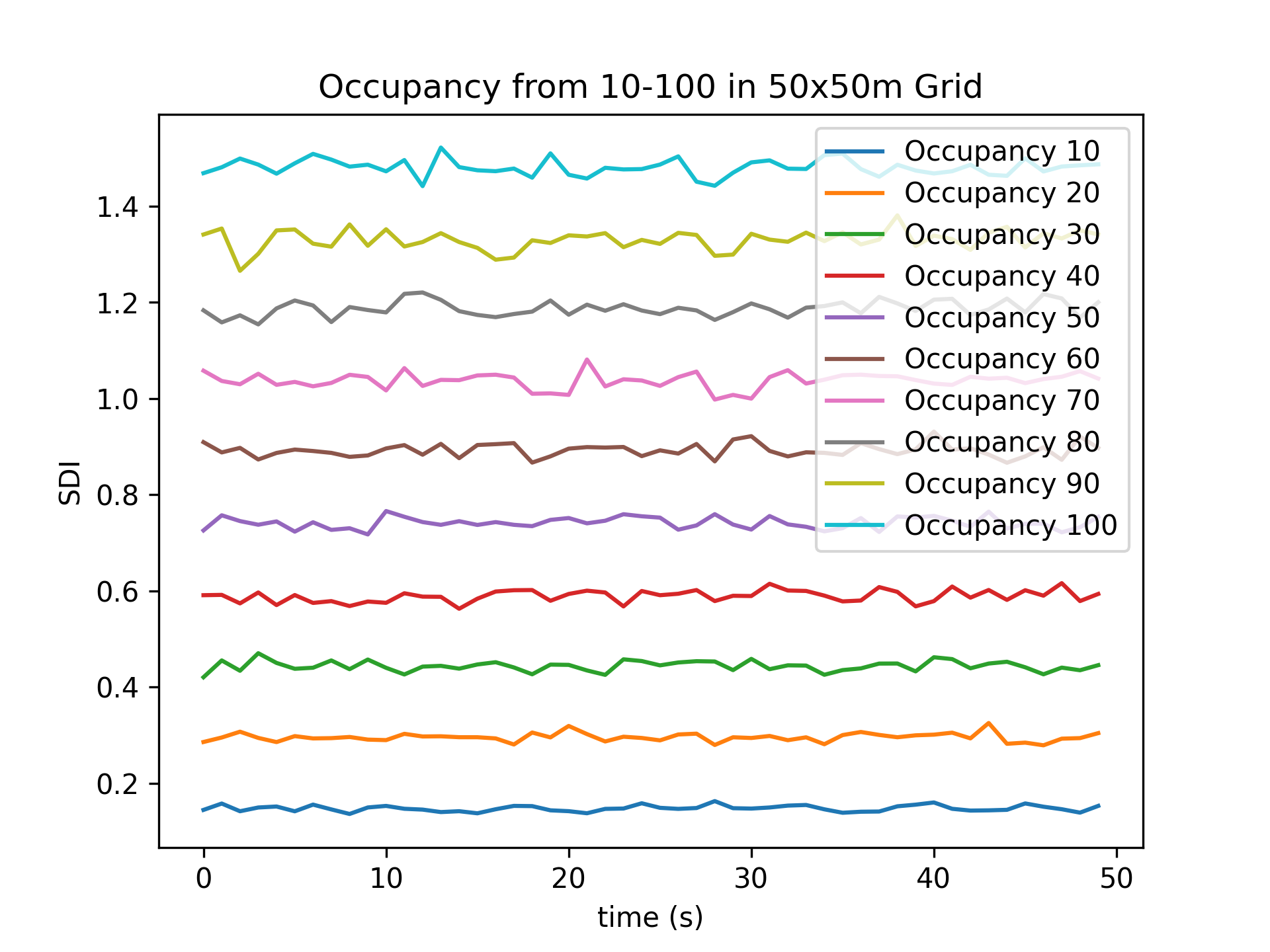}
    \caption{Increasing Occupancy}
    \label{fig:sdi_occ}
  \end{subfigure}    
  \begin{subfigure}[t]{.45\linewidth}
    \centering
    \includegraphics[width=\linewidth]{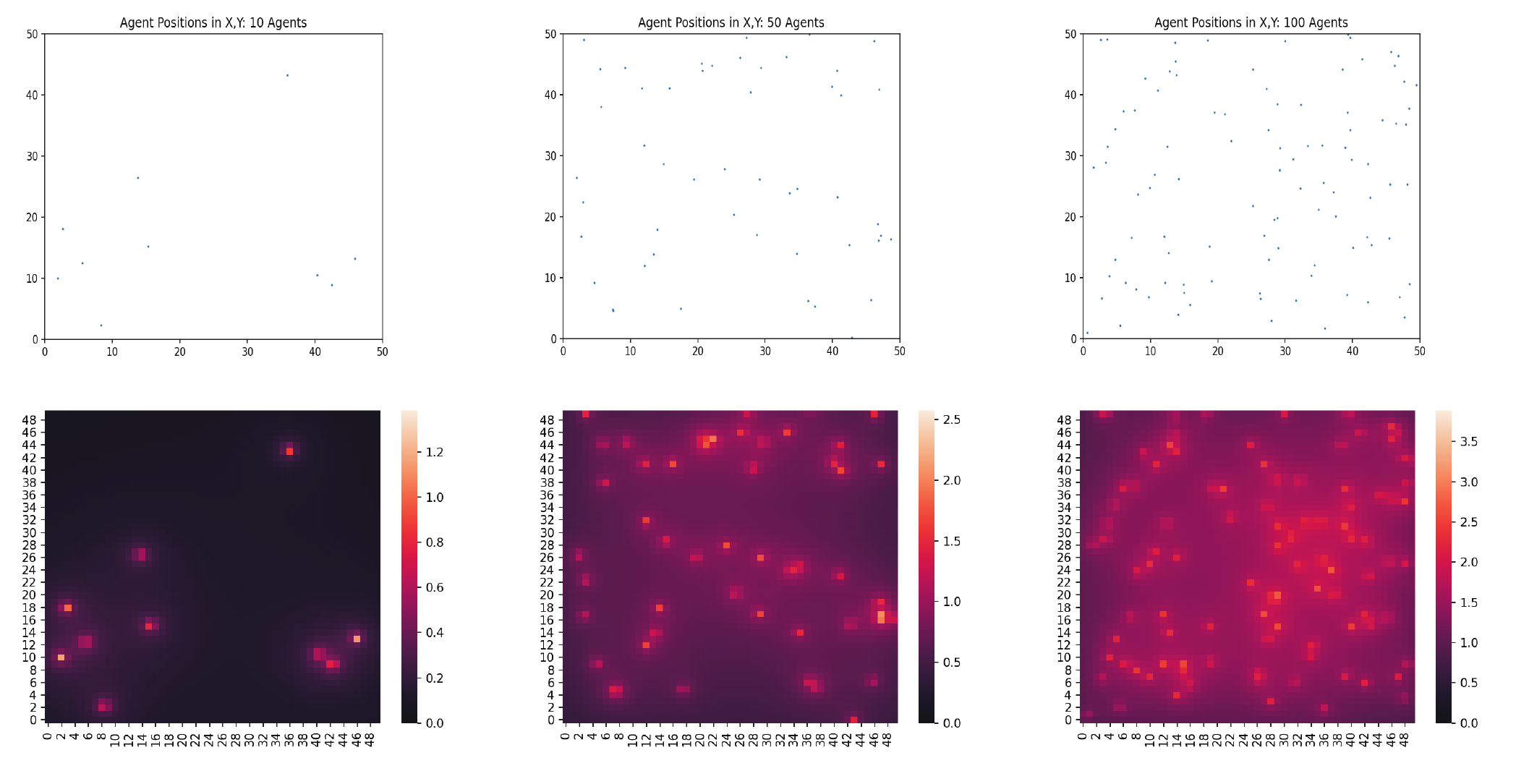}
    \caption{Heatmap of Occupancy for 10, 50, 100}
    \label{fig:sdi_heatmap}
  \end{subfigure}    
  
     \caption{We compute the \ac{SDI} on a $50\times50$ $m^2$ grid for easier visualizations on the impact of adding virtual agents to the environment.}
  \label{fig:sdi}
\end{figure*}

The \ac{SDI} is a calculation method that evaluates an environment's proclivity to facilitating distances among occupants with respect to both space and time. While air exchange is included indirectly, this is to provide an estimate and can be applicable to smell, not only human-human disease spread. It is a function of the distances occupants can maintain and the maximum occupancy of the environment. In the real world, an occupant is a person. However, as this calculation is done in simulation, we will refer to the representation of an occupant as an agent.

Given a virtual environment, we define $h$ as the air exchange factor, $m$ and $q$ respectively as the minimum and maximum distancing threshold, $T$ as the total time of simulation, and a set $A$ as the agents in a simulation. However not all agents are in the environment at a given time (e.g., at the beginning of the simulation), hence we define a subset $P_t \subset A$ for time $t$ where $P_t= \{ p \in A | p ~\text{is in the environment at time } t \}$. Matrix $\mathbf{C}$ contains the cells located in the environment. For a given cell, $\vec{c_{ij}}$ denotes the vector of scores, where the $i,j$ subscript corresponds to the $x,y$ center of the cell in the virtual environment, and the vector indices are ordered over time.

\begin{equation}
\vec{c_{ij}}(t) = \left(  \sum_{a \in P_t}^{} d_{ij}(a)  \right) \cdot h
\end{equation}

\noindent where $d_{ij}(a)$ is a function that computes a score relating the distance between a cell $\vec{c_{ij}}$ and an agent $a$ as follows:

\begin{equation}
d_{ij}(a) = \frac{m}{k(w)}
\end{equation}

\begin{equation*}
k(w) =
\begin{cases}
            m & \text{if } w < m \\
\texttt{none} & \text{if } w > q \\
            w & \text{otherwise} \\
\end{cases}
\end{equation*}

$$w = \texttt{dst}(c_{ij}, a)$$

To account for air quality, the air exchanges conducted naturally and by the HVAC system, a secondary factor $h$ can be applied to decrease the score over time.  $h$ is a factor from 0 to 1, based on the air exchange rate and time since the last simulation step, where no air exchange is 1 (the calculation does not change), to the ability to completely exchange the air instantly as 0. While this factor is motivated by research into pandemics, it is applicable to smell as well. 

Next, an environment score $E(t)$ is defined for each time $t$ as the mean of the cell scores.

\begin{equation}
  E(t)=\frac{1}{||\mathbf{C}||}\sum_{c_{i,j}\in\mathbf{C}}{\vec{c_{ij}}(t)}
\end{equation}

With the above approach, a general SDI metric for graph $g$ can be determined, such that the environment is scored by the average of all $E(t)$ at a given occupancy load $O_L$. 

$$ \texttt{SDI}(g|O_L) =  \frac{1}{T}\sum_{t=0}^{T}E(t)$$ 

However, the score of an environment should also consider the ability for the environment to maintain distancing at a rate that is at least the same, but preferably smaller, than the rate of the number of occupants. Therefore, we run the simulation with a varying number of agents over the time of the simulation, where the maximum number of agents at any time is defined by $O_{\text{max}}$.

In Figure \ref{fig:sdi_occ}, the increase in occupants by increments of 10 increases the \ac{SDI} proportionally. Intuitively, the more agents in the scene, the less distance can be maintained between them. Likewise, we show Figure~\ref{fig:sdi_heatmap}, with the top row showing agent positions and the bottom row showing the cell scores visualized as a heatmap (brighter is a higher \ac{SDI}).

While Figure~\ref{fig:sdi} demonstrates the impact of various parameters in the \ac{SDI}, for the optimization experiments, we use a large enough maximum distance $m$ of 1000m which maintains influence of all agents, a minimum distance $q$ of 0.3m representing the radius of an agent, and an environment exchange factor $h$ of 1.

\section{Optimization Framework} \label{optimization}

For each generation, after all the simulation sessions of a generation finish and all the corresponding \ac{SDI} values are calculated, one optimizer step is taken to decide and a new generation of graphs is produced, starting the next round of simulation. The optimization is terminated when a fixed number of steps has been taken, or based on a calculated threshold.

Figure \ref{tab:term} shows the symbols and notations used in this paper, and a complete pipeline is shown in Figure \ref{fig:optim_pipe}.

\subsection{Optimization Formulation}

Given the structural policy graph $g$ and a specified occupancy load parameter $O_L$, the optimization problem is formulated as
\[\argmin_{e_{ij}\in G} L(g|O_L)\text{ s.t. } \forall v_{i},v_{j}\in g,v_{j}\in R_{G}(v_i)\]
where $e_{ij}$ is an edge connecting nodes $v_i$ and $v_j$ in the structural graph $g$, and $L(g|O_L)$ can be an arbitrary scoring metric that measures the performance of the graph, e.g., $L(g)\coloneqq \texttt{SDI}(G|O_L)$. $R_G(v_i)$ is the set of all the strongly connected components of $v_i$.

In other words, our optimization process minimizes the target function by changing the state of the edges, while maintaining the strong connectivity of the structural graph. As an NP-optimization (NPO) problem, this process is an NP-complete combinatorial optimization that requires probabilistic methods (e.g., simulated annealing) to approach the optimal.

\subsection{Optimization Approach}

Due to the NP complexity with regard to the number of nodes in the graph, we use a mixed approach by combining GA and SA to search for the optimal policy. Our GA-SA approach is based on a few definitions and assumptions.

\subsubsection{Edit Distance}

\begin{figure}
  \centering
  \includegraphics[width=.6\linewidth]{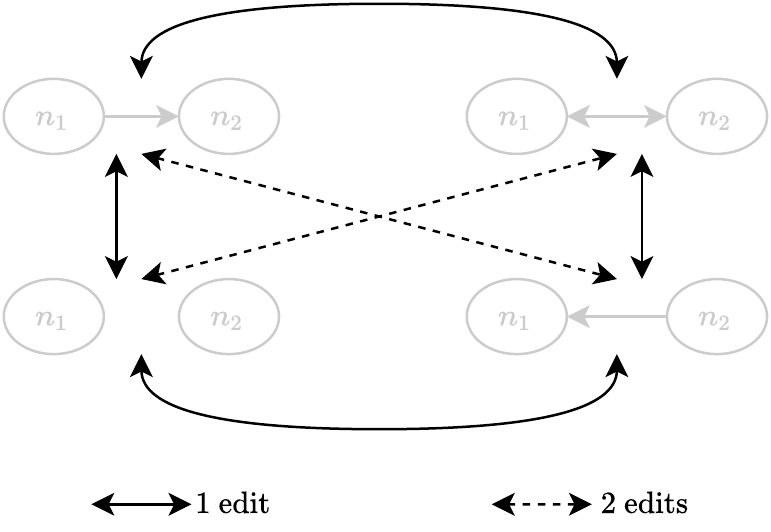}
  \caption{Definition of the edit distance for a pair of isomorphic edges. Edges with shorter edit distance are considered as more similar.}
  \label{fig:edit_dist}
\end{figure}

Given two isomorphic structural graphs $g_1$ and $g_2$, the edit distance $d(g_1,g_2)$ is defined as the total number of edge edits to make $g_1$ and $g_2$ identical. Formally,
$$d(g_1,g_2)=\sum_{e_{ij}\in g_1,e'_{ij}\in g_2}{d_e(e_{ij},e'_{ij})}.$$
Given two isomorphic edges, the edit distance value $d_e(e_{ij},e'_{ij})$ is defined by Figure \ref{fig:edit_dist}, which shows a total of 4 possible states of edges between nodes $n_1$ and $n_2$: (i) $n_1\rightarrow n_2$, (ii) $n_1\leftrightarrow n_2$, (iii) $n_1\quad n_2$, and (iv) $n_1\leftarrow n_2$. The solid lines mark 1 edit between states, while the broken lines mark 2 edits, e.g., the cost from state (i) to (ii) requires 1 edit, and that from state (i) to (iv) requires 2 edits.

\subsubsection{Similarity Monotonicity}

We assume that graphs with greater similarity yield closer \ac{SDI} values. Formally, given graphs $g,g_1,g_2$ and an arbitrary mapping function $f(g):g\rightarrow\mathbb{R}$, we have
$$d(g,g_1)\leq d(g,g_2)\Rightarrow|f(g)-f(g_1)|\leq|f(g)-f(g_2)|.$$
That said, given a graph $g$ and its corresponding score $s_g$, we will get larger score difference if we apply more edge edits on $g$. Note that we assume ``greater similarity'' is just sufficient to conclude ``smaller difference,'' so this assumption is not reversible: we don't assume that graphs are more similar \textit{due to} smaller \ac{SDI} difference.

\subsection{Algorithms}

Based on the formulation above, we propose this GA-SA method, described in the following algorithms. The pipeline of the optimization process is described in Algorithm \ref{alg:overview} and Figure \ref{fig:optim_pipe}, and detailed implementation of some functions are expanded in Algorithm \ref{alg:production}.

\begin{algorithm}
\small
  \SetKwInOut{Input}{input}
  \SetKwInOut{Output}{output}
  \SetAlgoNoEnd
  
  \Input{Structural graph $g_0$, Children count $n_c$, Edit distance $d$, Window size $w$, Optimization threshold $\hat{t}$, SA scalar $a$}
  
  $g\gets g_0$\tcp*[f]{Initial graph}\\
  $n\gets0$\tcp*[f]{Generation counter}\\
  $m_0\gets0$\tcp*[f]{Average SDI of most recent $w$ generations}\\
  $m_{-1}\gets0$\tcp*[f]{Average SDI of previous $w$ generations}\\
  $S\gets\emptyset$\tcp*[f]{Array of the final SDI of each generation}\\
  \While{$n<w$ \textbf{or} $n\geq w$ \textbf{and} $\frac{|m_0-m_{-1}|}{m_{-1}}>\hat{t}$}{
    $G\gets Produce(g,n_c,d)$\\
    \For{each $g'\in G$}{
      $h\gets Populate(g')$\tcp*[f]{A new navigational graph}\\
      Run simulation based on $h$\\
      Calculate \ac{SDI} of $g'$
    }
    \If{$Random(0,1)<a\exp(-\frac{1}{n})$}{
      $g\gets g'\in G$ with best \ac{SDI}\tcp*[f]{Conditional accepting}
    }
    $S\gets S\cup\{\text{ SDI of }g\}$\tcp*[f]{Save final SDI of generations}\\
    $n\gets n+1$\\
    \If{$n>w$}{
      $m_{-1}\gets m_0$\\
      $m_0\gets\frac{1}{w}\sum_{i=n-w}^{n}{s_i\in S}$\tcp*[f]{Average SDI}
    }
  }
\caption{Optimization iteration overview. We start with the initial graph $g_0$ and produce a generation of size $n_c$, then run the simulation with each child and get their corresponding \ac{SDI} values. We pick the best child as the parent for the next generation with a probability, and terminate the optimization process when the SDI value doesn't change significantly.}
\label{alg:overview}
\end{algorithm}

When starting the optimization process, we run the optimization process on the first $w$ generations of graphs. Starting from the $n$-th generation ($n\geq w+1$), we keep track of the mean value of the best \ac{SDI} scores of the recent $w$ generations and that of the $(n-1)$-th generation. When a generation is finished, we check if the value change is greater than a threshold and if so, the optimization process is considered ended.

As is shown in Figure \ref{fig:graphs}, the $Populate(\cdot)$ function processes the structural graph with the following steps:

\begin{enumerate}
  \item Inherit all the nodes in the structural graph to the navigational graph;
  \item Populate navigation nodes between the structural nodes with proper information, i.e., coordinates and item tags;
  \item Connect item nodes with joint nodes with edges complying to the policies defined in the corresponding structural graph.
\end{enumerate}

The navigational graphs are only used during the simulation and to calculate the scores of the graphs, and hence are discarded when the current iteration is over and will not be carried to the next generation.

The production function uses the GA method to generate new graphs in the next generation. In the $RandomizeState(\cdot)$ function, the state of the edge will mutate according to the transition relationship shown in Figure \ref{fig:edit_dist}. There are four states that the edges $e$ connecting nodes $n_1$ and $n_2$ can mutate to, and the edge must mutate to one of the rest 3 states where it is not currently in.

\begin{algorithm}
\small
  \SetKwInOut{Input}{input}
  \SetKwInOut{Output}{output}
  \SetAlgoNoEnd
  
  \Input{Base structural graph $g$, Sibling count $s$, Edit distance $d$}
  $n\gets 0$\\
  $G'\gets\emptyset$\\
  \While{$n\leq s$}{
    $n_e\gets0$\\
    $g'\gets\emptyset$\\
    \While{$n_e\leq d$}{
      Randomly pick one edge $e\in g$\\
      $e'\gets RandomizeState(e)$\tcp*[f]{Mutate edge}\\
      $n_e\gets n_e+d_e(e,e')$\\
      $g'\gets g'\cup\{e'\}$\\
    }
    \If{$g'\in G'$ \textbf{or} $g'$ is not strongly connected}{
      \textbf{continue}\\
    }
    $G'\gets G'\cup\{g'\}$\\
    $n\gets n+1$\\
  }
  \Return{$G'$}\\
    
\caption{Production function $Produce(\cdot)$. This function mutates the parent graph and generates all the children structural graphs for the next simulation sessions.}
\label{alg:production}
\end{algorithm}

When all the children from the current generation finish simulation and evaluation, all the SDI values are used by the optimizer to decide whether to accept the best-performed children as the parent for the next generation or reject it with a threshold and keep the original parent. The threshold decreases as the generation counter increase. This process uses the concept of conditional acceptance in the SA approach, as is shown in line 12--13 of algorithm \ref{alg:overview}, to prevent the algorithm from falling into a local optimum.

\section{Evaluation and Case Study}\label{sec:eval}

We evaluate our framework with 2 sections: (i) quantitative and qualitative analysis of \ac{SDI}, and (ii) simulation-optimization case studies: a simple one with the scoring metric basing on walking distance and another one using the \ac{SDI} metric.

\subsection{Distance-based Optimization}

In this scenario, the policy optimization process is based on the total traveling distance of all the agents, i.e., the optimizer aims to minimize the traveling distance when optimizing the policies. We verify our optimizer with 2 cases: one without the simulation but the accumulated Euclidean distance by traversing all the items in the shopping list, while the other using live simulations to collect the actual translation data of the agent model traveling in the 3D environment.

The experiments are conducted as follows. To calculate the Euclidean distance without running the simulation, we generate a random shopping list with $i$ shopping items, then accumulate the path length from the entrance to the 1st item, then from the 1st item to the 2nd, etc., eventually from the $(i-1)$-th item to the $i$-th item. Similarly, for the simulation-based experiment, the path length is calculated by sampling the positions of the agents by each frame.

\begin{figure}
  \centering
    \begin{subfigure}[h]{1.0\linewidth}
    \centering
        \begin{subfigure}[h]{0.5\linewidth} 
            \centering
            \includegraphics[width=\linewidth]{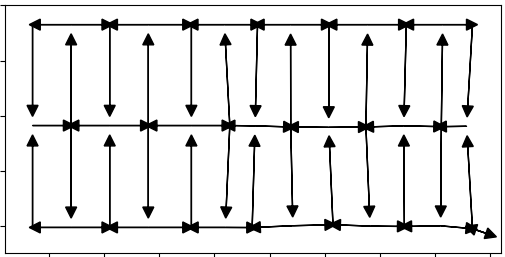}
            \caption{Initial graph}
            \label{subfig:init_dist}
        \end{subfigure}
    \end{subfigure}
    
    \begin{subfigure}[h]{1.0\linewidth}
      \begin{subfigure}[h]{0.49\linewidth} 
        \centering
        \includegraphics[width=\linewidth]{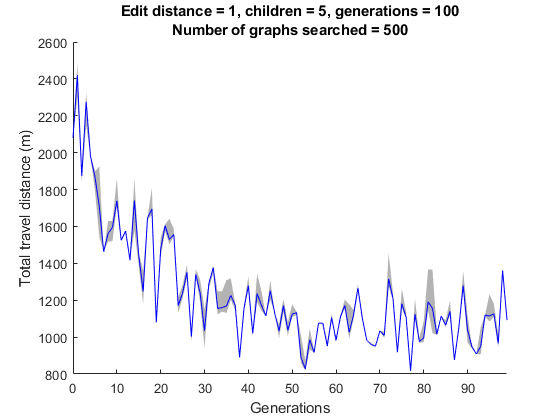}
        \caption{Static convergence curve}
        \label{subfig:opt_proc_th}
      \end{subfigure}
      \hfill
      \begin{subfigure}[h]{0.49\linewidth} 
            \centering
            \includegraphics[width=\linewidth]{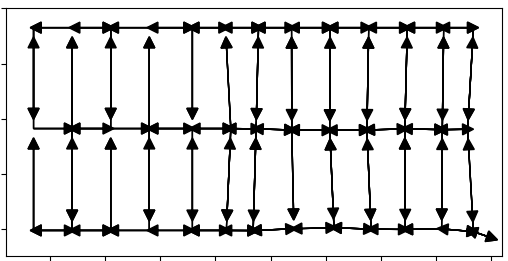}
            \caption{Static optimized graph}
            \label{subfig:opt_dist_th}
      \end{subfigure}
    \end{subfigure}
    
    \begin{subfigure}[h]{1.0\linewidth}
        \begin{subfigure}[h]{0.49\linewidth}
            \centering
            \includegraphics[width=\linewidth]{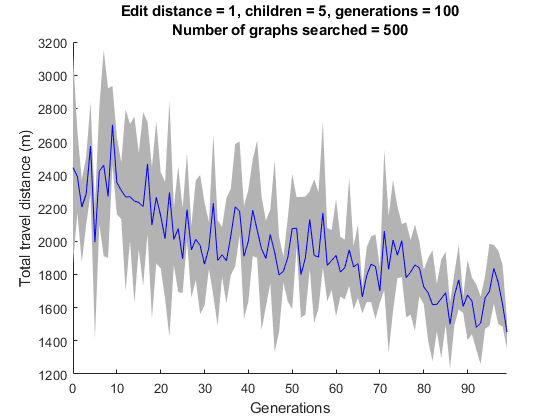}
            \caption{Simulated convergence curve}
            \label{subfig:opt_proc_sim}
        \end{subfigure}
        \hfill
        \begin{subfigure}[h]{0.49\linewidth}
            \centering
            \includegraphics[width=\linewidth]{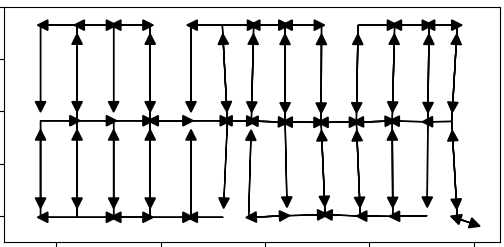}
            \caption{Simulated optimized graph}
            \label{subfig:opt_dist_sim}
        \end{subfigure}
    \end{subfigure}    
    
    \caption{\ref{subfig:init_dist} shows the initial policy graph with uni-directional edges for the optimization, \ref{subfig:opt_dist_th} shows an optimal policy optimized for static distance (theoretical), \ref{subfig:opt_dist_sim} shows an optimal policy optimized for simulated distance, \ref{subfig:opt_proc_th} shows the objective convergence for \ref{subfig:opt_dist_th}, and \ref{subfig:opt_proc_sim} shows the objective convergence for \ref{subfig:opt_dist_sim}.}
    \label{fig:unittest}
\end{figure}

The results in Figure \ref{fig:unittest} show that the optimization approach we used can be applied to the traveling distance data collected via simulation. The shaded part of the optimization curve shows the range of traveling distance of all the children in the current generation. As is shown in the graph, starting from a directional policy (\ref{subfig:init_dist}), both experiments reduce the total travel distance as the optimization proceeds. Both experiments show similar convergence rates, the variance of the simulation experiment is larger and the convergence speed is slower. A possible reason causing such variance is the simulation artifacts. The traveling distance is calculated by accumulating the translation of the agents by each frame, which is influenced by the position sampling interval, locomotion, and animation system. A higher sampling rate can help record the data with higher accuracy.

\subsection{Virtual Retail Environment}



\begin{figure*}[h!]
  \centering
  \begin{subfigure}[b]{0.55\linewidth}
      \begin{subfigure}[b]{0.44\linewidth}
        \centering
        \includegraphics[width=\linewidth]{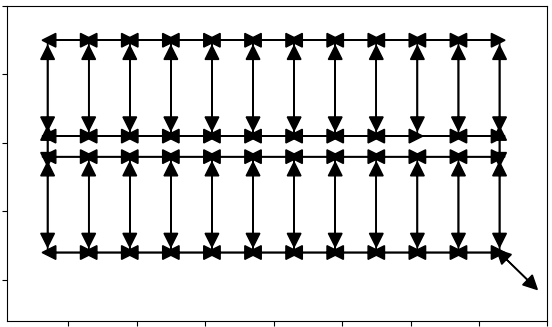}
        \caption{Beginning state of the graph}
        \label{subfig:begin_graph}
      \end{subfigure}
      \hfill
      \begin{subfigure}[b]{0.50\linewidth}
        \centering
        \includegraphics[width=\linewidth]{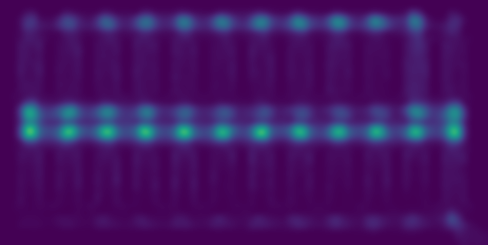}
        \caption{Initial heatmap (agent positions)}
        \label{subfig:begin_heat}
      \end{subfigure}
      
      \begin{subfigure}[b]{0.44\linewidth}
        \centering
        \includegraphics[width=\linewidth]{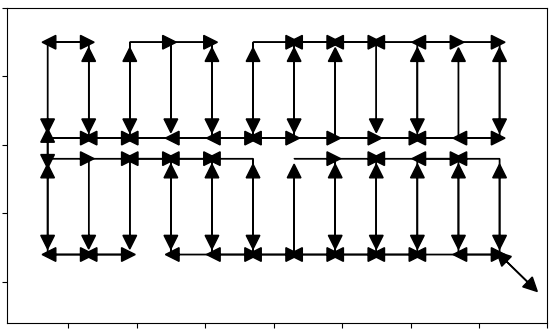}
        \caption{Optimized policy}
        \label{subfig:end_graph}
      \end{subfigure}
      \hfill
      \begin{subfigure}[b]{0.50\linewidth}
        \centering
        \includegraphics[width=\linewidth]{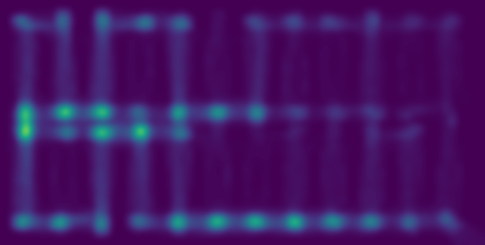}
        \caption{Final heatmap (agent positions)}
        \label{subfig:end_heat}
      \end{subfigure}
  \end{subfigure}
  \hfill
  \begin{subfigure}[b]{0.4\linewidth}
    \centering
    \includegraphics[width=\linewidth]{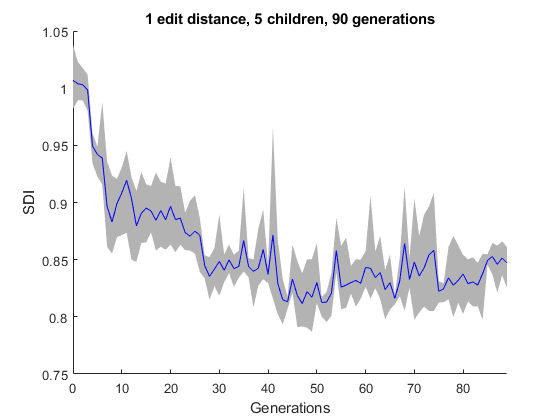}
    \caption{Optimization process}
    \label{subfig:exp_curve}
  \end{subfigure}
  \caption{The policy optimization example of a virtual retail store. \ac{SDI} is used as the optimization scoring metric. \ref{subfig:begin_graph} shows an initial policy graph, \ref{subfig:begin_heat} shows the starting heatmap of agent positions, \ref{subfig:end_graph} shows the optimized policy, \ref{subfig:end_heat} shows the ending heatmap of agent positions, and \ref{subfig:exp_curve} shows the objective convergence.}
  \label{fig:experiment_simple}
\end{figure*}

We created a virtual grocery store that resembles a typical configuration of a grocery market with vertically aligned shelves and horizontal hallways. Based on a constant occupancy setting, we conducted a complete simulation-optimization experiment to search for the optimal policy for this scenario. The optimization starts with the graph that consists of no directional constraints, as is shown in Figure \ref{subfig:begin_graph}. This is considered as the baseline case for a store environment: normally, all customers can move in the store freely in any direction. Apparently, this policy can potentially cause jamming and blockage, leading to a high \ac{SDI} value.

\begin{figure*}[h!]
  \begin{subfigure}[b]{0.33\linewidth}
    \centering
    \includegraphics[width=0.75\linewidth]{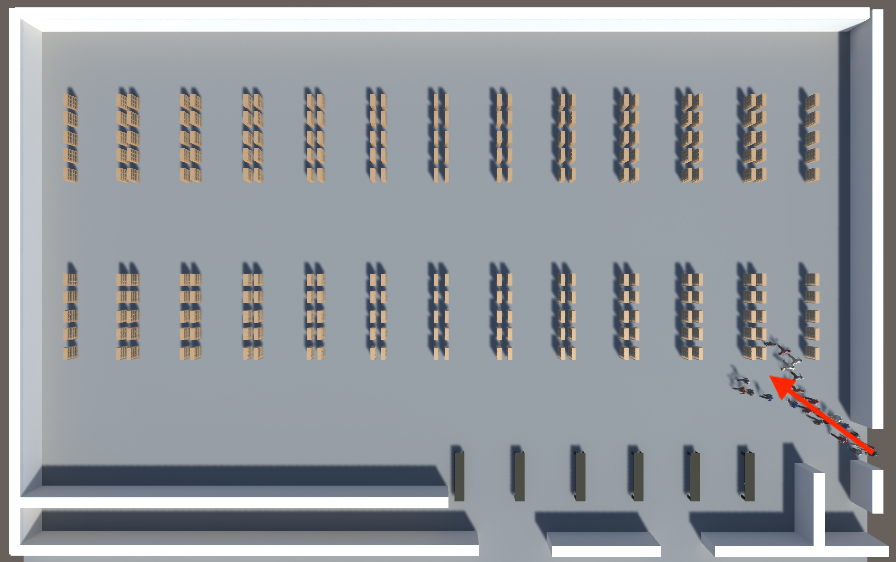}
  \end{subfigure}
  \hfill
  \begin{subfigure}[b]{0.33\linewidth}
    \centering
    \includegraphics[width=0.75\linewidth]{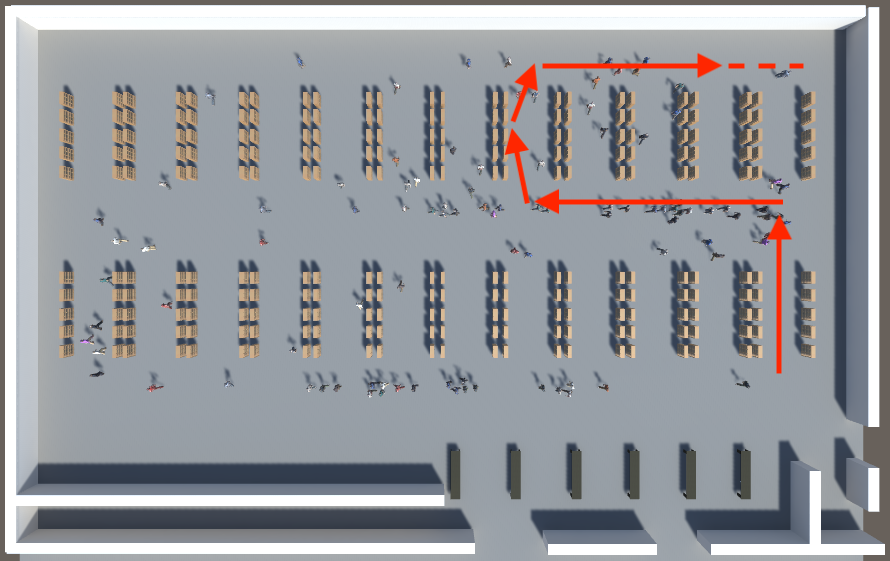}
  \end{subfigure}
  \hfill
  \begin{subfigure}[b]{0.33\linewidth}
    \centering
    \includegraphics[width=0.75\linewidth]{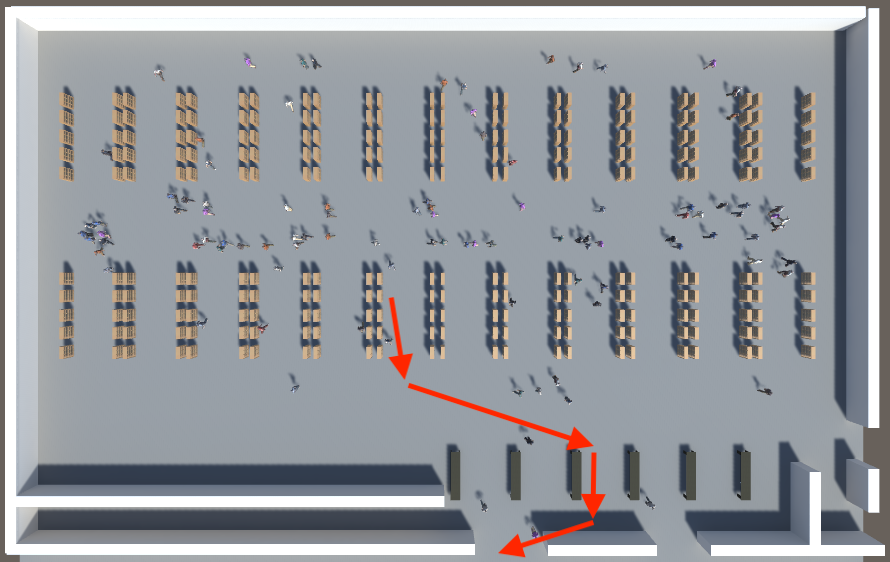}
  \end{subfigure}
  \caption{Simulation snapshots: (Left) In the beginning, agents are entering the facility, (Middle) During the simulation, agents are shopping the items following the policies, and (Right) At the end, agents are gradually finishing up the shopping and starting to queue up at the checkout counters.}
  \label{fig:sim_vis}
\end{figure*}

During the simulation, the environment is initially empty, and the agents start entering and conduct their shopping behaviors by visiting each target item shown in their shopping lists. When the occupancy load is reached, no more agents are allowed to enter, until any agent in the scene finishes the checkout behavior and exits the scene. Figure \ref{fig:sim_vis} shows 3 typical states of one simulation session: agent entering, shopping, and exiting.

The experiment uses 1 edit distance to produce the children of every generation, and the size of each generation is 5. The window size $w$ is 20 and the stopping threshold is $10^{-5}$. We run the simulation for a simulated time of 30 seconds of each scenario to collect data for the \ac{SDI} model, then the optimizer decides the graphs to use for the next iteration using the returned result of SDI. The optimization process finished in 93 generations. As is shown in Figure \ref{subfig:exp_curve}, The results show that our optimization process reduced \ac{SDI} by about 20\%. Similar to \ref{subfig:opt_dist_th} and \ref{subfig:opt_proc_th}, the shaded area shows the range of the \ac{SDI} values of all the children in a generation.

\section{Conclusion}\label{sec:conclusion}

In this paper we demonstrate how a graph optimization technique can be used on navigation graphs, in combination with a standard agent simulator, to find a policy that enables the largest spacing between virtual occupants during a shopping sequence.  The approach uses a combination of genetic algorithm and simulated annealing to change a \textit{directed} navigational graph that virtual occupants follow, while a distance-based metric records regions of densities in the environment. 

While we demonstrate a valid optimization through an intuitive and descriptive metric (distance traveled), the ability for the navigational policies to naturally influence occupant distances is more nuanced.  Specifically, if the \ac{SDI} is a properly weighted value, or how people would respond to such policies would be potential studies in future work to validate such approaches, for example through modifications to real buildings. Likewise, our simulation assumes some ideal conditions, such as agents strictly following the policies. Alternative behaviors could be implemented so that a probability of agents not following the policy may be useful. In future work, extending our framework with more robust agent-environment interaction could consider likely sources of disease spread, such as contaminated surfaces, requiring agents to physically interact with objects (e.g., picking up a box of cereal on a grocery shelf or opening the refrigerator door).  

It is important to consider if the \ac{SDI}, or any related work in the community, is meant as a method of tracking probabilities of infection rate, or instead taken more literally to be maintaining distance. While the work here is posed as an aid in the development of navigation policies for environments that have clear application to pandemic response, numerous issues at present limit this application in situ. Researches on social distancing measures aiding in preventing disease are often discussed in terms of pure isolation, or modeled in a way that assumes the number of people that have socially distanced are in fact not even in the same location (i.e., social distancing is treated as lockdown)~\cite{moosa2020effectiveness}
or are simply counting 'contacts' without reference to the distance or nature of it~\cite{zhang2020changes}. Likewise, researchers have emphasized the idea of social distancing set to 2 meters is outdated~\cite{jones2020two}. Therefore, simply recording the times two people are within that threshold may not be meaningful (i.e., recording 201 cm as safe and 199cm as dangerous). If the virus is transmitted in aerosols, interventions such as air quality control are required~\cite{lednicky2020viable}. Therefore, our work is not specifically targeting recommendations for the reduction of disease, but rather as a generalized model that could (1) optimize navigational policies for any metric and (2) a distance-based index that applies outside of pandemics as well.

\begin{acks}
This research has been partially funded by grants from ISSUM, and in 
part by NSF awards: IIS-1703883, IIS-1955404, and IIS-1955365.
\end{acks}
\bibliographystyle{ACM-Reference-Format}
\bibliography{simaud}


\begin{thebibliography}{39}


\ifx \showCODEN    \undefined \def \showCODEN     #1{\unskip}     \fi
\ifx \showDOI      \undefined \def \showDOI       #1{#1}\fi
\ifx \showISBNx    \undefined \def \showISBNx     #1{\unskip}     \fi
\ifx \showISBNxiii \undefined \def \showISBNxiii  #1{\unskip}     \fi
\ifx \showISSN     \undefined \def \showISSN      #1{\unskip}     \fi
\ifx \showLCCN     \undefined \def \showLCCN      #1{\unskip}     \fi
\ifx \shownote     \undefined \def \shownote      #1{#1}          \fi
\ifx \showarticletitle \undefined \def \showarticletitle #1{#1}   \fi
\ifx \showURL      \undefined \def \showURL       {\relax}        \fi
\providecommand\bibfield[2]{#2}
\providecommand\bibinfo[2]{#2}
\providecommand\natexlab[1]{#1}
\providecommand\showeprint[2][]{arXiv:#2}

\bibitem[\protect\citeauthoryear{AnyLogic}{AnyLogic}{[n.d.]}]%
        {anylogic}
AnyLogic \bibinfo{year}{[n.d.]}\natexlab{}.
\newblock \bibinfo{title}{{AnyLogic simulation software}}.
\newblock \bibinfo{howpublished}{\url{https://www.anylogic.com/}}.
\newblock
\newblock
\shownote{Accessed: 2020-12-02.}


\bibitem[\protect\citeauthoryear{Banerjee and Kraemer}{Banerjee and
  Kraemer}{2010}]%
        {banerjee2010evaluation}
\bibfield{author}{\bibinfo{person}{Bikramjit Banerjee} {and}
  \bibinfo{person}{Landon Kraemer}.} \bibinfo{year}{2010}\natexlab{}.
\newblock \showarticletitle{Evaluation and comparison of multi-agent based
  crowd simulation systems}. In \bibinfo{booktitle}{\emph{Workshop on Agents
  for Games and Simulations}}. Springer, \bibinfo{pages}{53--66}.
\newblock


\bibitem[\protect\citeauthoryear{Berseth, Usman, Haworth, Kapadia, and
  Faloutsos}{Berseth et~al\mbox{.}}{2015}]%
        {berseth2015environment}
\bibfield{author}{\bibinfo{person}{Glen Berseth}, \bibinfo{person}{Muhammad
  Usman}, \bibinfo{person}{Brandon Haworth}, \bibinfo{person}{Mubbasir
  Kapadia}, {and} \bibinfo{person}{Petros Faloutsos}.}
  \bibinfo{year}{2015}\natexlab{}.
\newblock \showarticletitle{Environment optimization for crowd evacuation}.
\newblock \bibinfo{journal}{\emph{Computer Animation and Virtual Worlds}}
  \bibinfo{volume}{26}, \bibinfo{number}{3-4} (\bibinfo{year}{2015}),
  \bibinfo{pages}{377--386}.
\newblock


\bibitem[\protect\citeauthoryear{Cassol, Testa, Jung, Usman, Faloutsos,
  Berseth, Kapadia, Badler, and Musse}{Cassol et~al\mbox{.}}{2017}]%
        {cassol2017evaluating}
\bibfield{author}{\bibinfo{person}{Vincius~J Cassol},
  \bibinfo{person}{Est{\^e}v{\~a}o~Smania Testa},
  \bibinfo{person}{Cl{\'a}udio~Rosito Jung}, \bibinfo{person}{Muhammad Usman},
  \bibinfo{person}{Petros Faloutsos}, \bibinfo{person}{Glen Berseth},
  \bibinfo{person}{Mubbasir Kapadia}, \bibinfo{person}{Norman~I Badler}, {and}
  \bibinfo{person}{Soraia~Raupp Musse}.} \bibinfo{year}{2017}\natexlab{}.
\newblock \showarticletitle{Evaluating and optimizing evacuation plans for
  crowd egress}.
\newblock \bibinfo{journal}{\emph{CGA}} \bibinfo{volume}{37},
  \bibinfo{number}{4} (\bibinfo{year}{2017}), \bibinfo{pages}{60--71}.
\newblock


\bibitem[\protect\citeauthoryear{Chirkin and K{\"o}nig}{Chirkin and
  K{\"o}nig}{2016}]%
        {chirkin2016concept}
\bibfield{author}{\bibinfo{person}{Artem~M Chirkin} {and}
  \bibinfo{person}{Reinhard K{\"o}nig}.} \bibinfo{year}{2016}\natexlab{}.
\newblock \showarticletitle{Concept of interactive machine learning in urban
  design problems}.
\newblock In \bibinfo{booktitle}{\emph{Proceedings of the Smart Cities for
  Better Living with HCI and UX}}. \bibinfo{pages}{10--13}.
\newblock


\bibitem[\protect\citeauthoryear{Coibion, Gorodnichenko, and Weber}{Coibion
  et~al\mbox{.}}{2020}]%
        {coibion2020cost}
\bibfield{author}{\bibinfo{person}{Olivier Coibion}, \bibinfo{person}{Yuriy
  Gorodnichenko}, {and} \bibinfo{person}{Michael Weber}.}
  \bibinfo{year}{2020}\natexlab{}.
\newblock \bibinfo{booktitle}{\emph{The cost of the covid-19 crisis: Lockdowns,
  macroeconomic expectations, and consumer spending}}.
\newblock \bibinfo{type}{{T}echnical {R}eport}. \bibinfo{institution}{National
  Bureau of Economic Research}.
\newblock


\bibitem[\protect\citeauthoryear{Das, Day, Hauck, Haymaker, and Davis}{Das
  et~al\mbox{.}}{2016}]%
        {das2016space}
\bibfield{author}{\bibinfo{person}{Subhajit Das}, \bibinfo{person}{Colin Day},
  \bibinfo{person}{John Hauck}, \bibinfo{person}{John Haymaker}, {and}
  \bibinfo{person}{Diana Davis}.} \bibinfo{year}{2016}\natexlab{}.
\newblock \showarticletitle{Space plan generator: Rapid generationn \&
  evaluation of floor plan design options to inform decision making}.
\newblock  (\bibinfo{year}{2016}).
\newblock


\bibitem[\protect\citeauthoryear{Dubey, Khoo, Morad, H{\"o}lscher, and
  Kapadia}{Dubey et~al\mbox{.}}{2020}]%
        {dubey2020autosign}
\bibfield{author}{\bibinfo{person}{Rohit~K Dubey}, \bibinfo{person}{Wei~Ping
  Khoo}, \bibinfo{person}{Michal~Gath Morad}, \bibinfo{person}{Christoph
  H{\"o}lscher}, {and} \bibinfo{person}{Mubbasir Kapadia}.}
  \bibinfo{year}{2020}\natexlab{}.
\newblock \showarticletitle{AUTOSIGN: A multi-criteria optimization approach to
  computer aided design of signage layouts in complex buildings}.
\newblock \bibinfo{journal}{\emph{Computers \& Graphics}}
  (\bibinfo{year}{2020}).
\newblock


\bibitem[\protect\citeauthoryear{Feng, Yu, Yeung, Yin, and Zhou}{Feng
  et~al\mbox{.}}{2016}]%
        {feng2016crowd}
\bibfield{author}{\bibinfo{person}{Tian Feng}, \bibinfo{person}{Lap-Fai Yu},
  \bibinfo{person}{Sai-Kit Yeung}, \bibinfo{person}{KangKang Yin}, {and}
  \bibinfo{person}{Kun Zhou}.} \bibinfo{year}{2016}\natexlab{}.
\newblock \showarticletitle{Crowd-driven mid-scale layout design.}
\newblock \bibinfo{journal}{\emph{Transaction on Graphics}}
  \bibinfo{volume}{35}, \bibinfo{number}{4} (\bibinfo{year}{2016}),
  \bibinfo{pages}{132--1}.
\newblock


\bibitem[\protect\citeauthoryear{Fort, Crespi, Elion, Kermanizadeh, Wani, and
  Lange}{Fort et~al\mbox{.}}{[n.d.]}]%
        {unitysim}
\bibfield{author}{\bibinfo{person}{James Fort}, \bibinfo{person}{Adam Crespi},
  \bibinfo{person}{Chris Elion}, \bibinfo{person}{Rambod Kermanizadeh},
  \bibinfo{person}{Priyesh Wani}, {and} \bibinfo{person}{Danny Lange}.}
  \bibinfo{year}{[n.d.]}\natexlab{}.
\newblock \bibinfo{title}{Exploring new ways to simulate the coronavirus
  spread}.
\newblock
  \bibinfo{howpublished}{\url{https://blogs.unity3d.com/2020/05/08/exploring-new-ways-to-simulate-the-coronavirus-spread/}}.
\newblock


\bibitem[\protect\citeauthoryear{Frank, Sallis, Saelens, Leary, Cain, Conway,
  and Hess}{Frank et~al\mbox{.}}{2010}]%
        {frank2010development}
\bibfield{author}{\bibinfo{person}{Lawrence~D Frank}, \bibinfo{person}{James~F
  Sallis}, \bibinfo{person}{Brian~E Saelens}, \bibinfo{person}{Lauren Leary},
  \bibinfo{person}{Kelli Cain}, \bibinfo{person}{Terry~L Conway}, {and}
  \bibinfo{person}{Paul~M Hess}.} \bibinfo{year}{2010}\natexlab{}.
\newblock \showarticletitle{The development of a walkability index: application
  to the Neighborhood Quality of Life Study}.
\newblock \bibinfo{journal}{\emph{Sports Medicine}} \bibinfo{volume}{44},
  \bibinfo{number}{13} (\bibinfo{year}{2010}), \bibinfo{pages}{924--933}.
\newblock


\bibitem[\protect\citeauthoryear{Hall, Birdwhistell, Bock, Bohannan,
  Diebold~Jr, Durbin, Edmonson, Fischer, Hymes, Kimball, et~al\mbox{.}}{Hall
  et~al\mbox{.}}{1968}]%
        {hall1968proxemics}
\bibfield{author}{\bibinfo{person}{Edward~T Hall}, \bibinfo{person}{Ray~L
  Birdwhistell}, \bibinfo{person}{Bernhard Bock}, \bibinfo{person}{Paul
  Bohannan}, \bibinfo{person}{A~Richard Diebold~Jr}, \bibinfo{person}{Marshall
  Durbin}, \bibinfo{person}{Munro~S Edmonson}, \bibinfo{person}{JL Fischer},
  \bibinfo{person}{Dell Hymes}, \bibinfo{person}{Solon~T Kimball},
  {et~al\mbox{.}}} \bibinfo{year}{1968}\natexlab{}.
\newblock \showarticletitle{Proxemics [and comments and replies]}.
\newblock \bibinfo{journal}{\emph{Current Anthropology}} \bibinfo{volume}{9},
  \bibinfo{number}{2/3} (\bibinfo{year}{1968}), \bibinfo{pages}{83--108}.
\newblock


\bibitem[\protect\citeauthoryear{Helbing, Farkas, and Vicsek}{Helbing
  et~al\mbox{.}}{2002}]%
        {helbing2002crowd}
\bibfield{author}{\bibinfo{person}{Dirk Helbing}, \bibinfo{person}{Ill{\'e}s~J
  Farkas}, {and} \bibinfo{person}{Tam{\'a}s Vicsek}.}
  \bibinfo{year}{2002}\natexlab{}.
\newblock \showarticletitle{Crowd disasters and simulation of panic
  situations}.
\newblock In \bibinfo{booktitle}{\emph{The Science of Disasters}}.
  \bibinfo{publisher}{Springer}, \bibinfo{pages}{330--350}.
\newblock


\bibitem[\protect\citeauthoryear{Helbing and Johansson}{Helbing and
  Johansson}{2013}]%
        {helbing2013pedestrian}
\bibfield{author}{\bibinfo{person}{Dirk Helbing} {and} \bibinfo{person}{Anders
  Johansson}.} \bibinfo{year}{2013}\natexlab{}.
\newblock \showarticletitle{Pedestrian, crowd, and evacuation dynamics}.
\newblock \bibinfo{journal}{\emph{arXiv preprint arXiv:1309.1609}}
  (\bibinfo{year}{2013}).
\newblock


\bibitem[\protect\citeauthoryear{Huang, Lin, Barrett, Springer, Wang, Pomplun,
  and Yu}{Huang et~al\mbox{.}}{2017}]%
        {huang2017automatic}
\bibfield{author}{\bibinfo{person}{Haikun Huang}, \bibinfo{person}{Ni-Ching
  Lin}, \bibinfo{person}{Lorenzo Barrett}, \bibinfo{person}{Darian Springer},
  \bibinfo{person}{Hsueh-Cheng Wang}, \bibinfo{person}{Marc Pomplun}, {and}
  \bibinfo{person}{Lap-Fai Yu}.} \bibinfo{year}{2017}\natexlab{}.
\newblock \showarticletitle{Automatic optimization of wayfinding design}.
\newblock \bibinfo{journal}{\emph{Transactions on Visualization and Computer
  Graphics}} \bibinfo{volume}{24}, \bibinfo{number}{9} (\bibinfo{year}{2017}),
  \bibinfo{pages}{2516--2530}.
\newblock


\bibitem[\protect\citeauthoryear{Johnson, Aragon, McGeoch, and Schevon}{Johnson
  et~al\mbox{.}}{1989}]%
        {johnson1989optimization}
\bibfield{author}{\bibinfo{person}{David~S Johnson}, \bibinfo{person}{Cecilia~R
  Aragon}, \bibinfo{person}{Lyle~A McGeoch}, {and} \bibinfo{person}{Catherine
  Schevon}.} \bibinfo{year}{1989}\natexlab{}.
\newblock \showarticletitle{Optimization by simulated annealing: An
  experimental evaluation; part I, graph partitioning}.
\newblock \bibinfo{journal}{\emph{Operations Research}} \bibinfo{volume}{37},
  \bibinfo{number}{6} (\bibinfo{year}{1989}), \bibinfo{pages}{865--892}.
\newblock


\bibitem[\protect\citeauthoryear{Jones, Qureshi, Temple, Larwood, Greenhalgh,
  and Bourouiba}{Jones et~al\mbox{.}}{2020}]%
        {jones2020two}
\bibfield{author}{\bibinfo{person}{Nicholas~R Jones}, \bibinfo{person}{Zeshan~U
  Qureshi}, \bibinfo{person}{Robert~J Temple}, \bibinfo{person}{Jessica~PJ
  Larwood}, \bibinfo{person}{Trisha Greenhalgh}, {and} \bibinfo{person}{Lydia
  Bourouiba}.} \bibinfo{year}{2020}\natexlab{}.
\newblock \showarticletitle{Two metres or one: what is the evidence for
  physical distancing in covid-19?}
\newblock \bibinfo{journal}{\emph{BMJ}}  \bibinfo{volume}{370}
  (\bibinfo{year}{2020}).
\newblock


\bibitem[\protect\citeauthoryear{Kapadia, Marshak, and Badler}{Kapadia
  et~al\mbox{.}}{2014}]%
        {kapadia2014adapt}
\bibfield{author}{\bibinfo{person}{Mubbasir Kapadia}, \bibinfo{person}{Nathan
  Marshak}, {and} \bibinfo{person}{Norman~I Badler}.}
  \bibinfo{year}{2014}\natexlab{}.
\newblock \showarticletitle{ADAPT: The agent development and prototyping
  testbed}.
\newblock \bibinfo{journal}{\emph{Transactions on Visualization \& Computer
  Graphics}} \bibinfo{number}{1} (\bibinfo{year}{2014}), \bibinfo{pages}{1}.
\newblock


\bibitem[\protect\citeauthoryear{Lednicky, Lauzard, Fan, Jutla, Tilly, Gangwar,
  Usmani, Shankar, Mohamed, Eiguren-Fernandez, et~al\mbox{.}}{Lednicky
  et~al\mbox{.}}{2020}]%
        {lednicky2020viable}
\bibfield{author}{\bibinfo{person}{John~A Lednicky}, \bibinfo{person}{Michael
  Lauzard}, \bibinfo{person}{Z~Hugh Fan}, \bibinfo{person}{Antarpreet Jutla},
  \bibinfo{person}{Trevor~B Tilly}, \bibinfo{person}{Mayank Gangwar},
  \bibinfo{person}{Moiz Usmani}, \bibinfo{person}{Sripriya~Nannu Shankar},
  \bibinfo{person}{Karim Mohamed}, \bibinfo{person}{Arantza Eiguren-Fernandez},
  {et~al\mbox{.}}} \bibinfo{year}{2020}\natexlab{}.
\newblock \showarticletitle{Viable SARS-CoV-2 in the air of a hospital room
  with COVID-19 patients}.
\newblock \bibinfo{journal}{\emph{Infectious Diseases}}  \bibinfo{volume}{100}
  (\bibinfo{year}{2020}), \bibinfo{pages}{476--482}.
\newblock


\bibitem[\protect\citeauthoryear{Miao, Koenig, and Knecht}{Miao
  et~al\mbox{.}}{2020}]%
        {miao2020development}
\bibfield{author}{\bibinfo{person}{Yufan Miao}, \bibinfo{person}{Reinhard
  Koenig}, {and} \bibinfo{person}{Katja Knecht}.}
  \bibinfo{year}{2020}\natexlab{}.
\newblock \showarticletitle{The Development of Optimization Methods in
  Generative Urban Design: A Review}. In \bibinfo{booktitle}{\emph{Proceedings
  of SimAUD}}.
\newblock


\bibitem[\protect\citeauthoryear{Moosa}{Moosa}{2020}]%
        {moosa2020effectiveness}
\bibfield{author}{\bibinfo{person}{Imad~A Moosa}.}
  \bibinfo{year}{2020}\natexlab{}.
\newblock \showarticletitle{The effectiveness of social distancing in
  containing Covid-19}.
\newblock \bibinfo{journal}{\emph{Applied Economics}} (\bibinfo{year}{2020}),
  \bibinfo{pages}{1--14}.
\newblock


\bibitem[\protect\citeauthoryear{Morawska, Tang, Bahnfleth, Bluyssen, Boerstra,
  Buonanno, Cao, Dancer, Floto, Franchimon, Haworth, Hogeling, Isaxon, Jimenez,
  Kurnitski, Li, Loomans, Marks, Marr, Mazzarella, Melikov, Miller, Milton,
  Nazaroff, Nielsen, Noakes, Peccia, Querol, Sekhar, Seppänen, ichi Tanabe,
  Tellier, Tham, Wargocki, Wierzbicka, and Yao}{Morawska et~al\mbox{.}}{2020}]%
        {MORAWSKA2020105832}
\bibfield{author}{\bibinfo{person}{Lidia Morawska}, \bibinfo{person}{Julian~W.
  Tang}, \bibinfo{person}{William Bahnfleth}, \bibinfo{person}{Philomena~M.
  Bluyssen}, \bibinfo{person}{Atze Boerstra}, \bibinfo{person}{Giorgio
  Buonanno}, \bibinfo{person}{Junji Cao}, \bibinfo{person}{Stephanie Dancer},
  \bibinfo{person}{Andres Floto}, \bibinfo{person}{Francesco Franchimon},
  \bibinfo{person}{Charles Haworth}, \bibinfo{person}{Jaap Hogeling},
  \bibinfo{person}{Christina Isaxon}, \bibinfo{person}{Jose~L. Jimenez},
  \bibinfo{person}{Jarek Kurnitski}, \bibinfo{person}{Yuguo Li},
  \bibinfo{person}{Marcel Loomans}, \bibinfo{person}{Guy Marks},
  \bibinfo{person}{Linsey~C. Marr}, \bibinfo{person}{Livio Mazzarella},
  \bibinfo{person}{Arsen~Krikor Melikov}, \bibinfo{person}{Shelly Miller},
  \bibinfo{person}{Donald~K. Milton}, \bibinfo{person}{William Nazaroff},
  \bibinfo{person}{Peter~V. Nielsen}, \bibinfo{person}{Catherine Noakes},
  \bibinfo{person}{Jordan Peccia}, \bibinfo{person}{Xavier Querol},
  \bibinfo{person}{Chandra Sekhar}, \bibinfo{person}{Olli Seppänen},
  \bibinfo{person}{Shin ichi Tanabe}, \bibinfo{person}{Raymond Tellier},
  \bibinfo{person}{Kwok~Wai Tham}, \bibinfo{person}{Pawel Wargocki},
  \bibinfo{person}{Aneta Wierzbicka}, {and} \bibinfo{person}{Maosheng Yao}.}
  \bibinfo{year}{2020}\natexlab{}.
\newblock \showarticletitle{How can airborne transmission of COVID-19 indoors
  be minimised?}
\newblock \bibinfo{journal}{\emph{Environment International}}
  \bibinfo{volume}{142} (\bibinfo{year}{2020}), \bibinfo{pages}{105832}.
\newblock
\showISSN{0160-4120}
\urldef\tempurl%
\url{https://doi.org/10.1016/j.envint.2020.105832}
\showDOI{\tempurl}


\bibitem[\protect\citeauthoryear{Nagy, Lau, Locke, Stoddart, Villaggi, Wang,
  Zhao, and Benjamin}{Nagy et~al\mbox{.}}{2017}]%
        {nagy2017project}
\bibfield{author}{\bibinfo{person}{Danil Nagy}, \bibinfo{person}{Damon Lau},
  \bibinfo{person}{John Locke}, \bibinfo{person}{Jim Stoddart},
  \bibinfo{person}{Lorenzo Villaggi}, \bibinfo{person}{Ray Wang},
  \bibinfo{person}{Dale Zhao}, {and} \bibinfo{person}{David Benjamin}.}
  \bibinfo{year}{2017}\natexlab{}.
\newblock \showarticletitle{Project Discover: An application of generative
  design for architectural space planning}. In
  \bibinfo{booktitle}{\emph{Proceedings of SimAUD}}. \bibinfo{pages}{1--8}.
\newblock


\bibitem[\protect\citeauthoryear{Nagy, Villaggi, and Benjamin}{Nagy
  et~al\mbox{.}}{2018}]%
        {nagy2018generative}
\bibfield{author}{\bibinfo{person}{Danil Nagy}, \bibinfo{person}{Lorenzo
  Villaggi}, {and} \bibinfo{person}{David Benjamin}.}
  \bibinfo{year}{2018}\natexlab{}.
\newblock \showarticletitle{Generative urban design: integrating financial and
  energy goals for automated neighborhood layout}. In
  \bibinfo{booktitle}{\emph{Proceedings of the Symposium for Architecture and
  Urban Design Design, Delft, the Netherlands}}. \bibinfo{pages}{265--274}.
\newblock


\bibitem[\protect\citeauthoryear{{Oasys}}{{Oasys}}{[n.d.]}]%
        {massmotion}
\bibfield{author}{\bibinfo{person}{{Oasys}}.}
  \bibinfo{year}{[n.d.]}\natexlab{}.
\newblock \bibinfo{booktitle}{\emph{MassMotion}}.
\newblock
\urldef\tempurl%
\url{https://www.oasys-software.com/products/pedestrian-simulation/massmotion/}
\showURL{%
\tempurl}


\bibitem[\protect\citeauthoryear{Organization}{Organization}{[n.d.]}]%
        {WHOsituation}
\bibfield{author}{\bibinfo{person}{World~Health Organization}.}
  \bibinfo{year}{[n.d.]}\natexlab{}.
\newblock \bibinfo{title}{Coronavirus disease (COVID-2019) situation reports}.
\newblock
  \bibinfo{howpublished}{\url{https://www.who.int/emergencies/diseases/novel-coronavirus-2019/situation-reports}}.
\newblock


\bibitem[\protect\citeauthoryear{Pettr{\'e}, Grillon, and Thalmann}{Pettr{\'e}
  et~al\mbox{.}}{2008}]%
        {pettre2008crowds}
\bibfield{author}{\bibinfo{person}{Julien Pettr{\'e}}, \bibinfo{person}{Helena
  Grillon}, {and} \bibinfo{person}{Daniel Thalmann}.}
  \bibinfo{year}{2008}\natexlab{}.
\newblock \showarticletitle{Crowds of moving objects: Navigation planning and
  simulation}.
\newblock In \bibinfo{booktitle}{\emph{SIGGRAPH 2008 classes}}.
  \bibinfo{pages}{1--7}.
\newblock


\bibitem[\protect\citeauthoryear{Reynolds}{Reynolds}{1999}]%
        {reynolds1999steering}
\bibfield{author}{\bibinfo{person}{Craig~W Reynolds}.}
  \bibinfo{year}{1999}\natexlab{}.
\newblock \showarticletitle{Steering behaviors for autonomous characters}. In
  \bibinfo{booktitle}{\emph{Game Developers Conference}},
  Vol.~\bibinfo{volume}{1999}. Citeseer, \bibinfo{pages}{763--782}.
\newblock


\bibitem[\protect\citeauthoryear{Schaumann, Breslav, Goldstein, Khan, and
  Kalay}{Schaumann et~al\mbox{.}}{2017}]%
        {schaumann2017simulating}
\bibfield{author}{\bibinfo{person}{Davide Schaumann}, \bibinfo{person}{Simon
  Breslav}, \bibinfo{person}{Rhys Goldstein}, \bibinfo{person}{Azam Khan},
  {and} \bibinfo{person}{Yehuda~E Kalay}.} \bibinfo{year}{2017}\natexlab{}.
\newblock \showarticletitle{Simulating use scenarios in hospitals using
  multi-agent narratives}.
\newblock \bibinfo{journal}{\emph{Building Performance Simulation}}
  \bibinfo{volume}{10}, \bibinfo{number}{5-6} (\bibinfo{year}{2017}),
  \bibinfo{pages}{636--652}.
\newblock


\bibitem[\protect\citeauthoryear{Shin and Lee}{Shin and Lee}{2019}]%
        {shin2019indoor}
\bibfield{author}{\bibinfo{person}{Jaeyoung Shin} {and}
  \bibinfo{person}{Jin-Kook Lee}.} \bibinfo{year}{2019}\natexlab{}.
\newblock \showarticletitle{Indoor Walkability Index: BIM-enabled approach to
  Quantifying building circulation}.
\newblock \bibinfo{journal}{\emph{Automation in Construction}}
  \bibinfo{volume}{106} (\bibinfo{year}{2019}), \bibinfo{pages}{102845}.
\newblock


\bibitem[\protect\citeauthoryear{Shoulson, Garcia, Jones, Mead, and
  Badler}{Shoulson et~al\mbox{.}}{2011}]%
        {shoulson2011parameterizing}
\bibfield{author}{\bibinfo{person}{Alexander Shoulson},
  \bibinfo{person}{Francisco~M Garcia}, \bibinfo{person}{Matthew Jones},
  \bibinfo{person}{Robert Mead}, {and} \bibinfo{person}{Norman~I Badler}.}
  \bibinfo{year}{2011}\natexlab{}.
\newblock \showarticletitle{Parameterizing behavior trees}. In
  \bibinfo{booktitle}{\emph{Motion in Games}}. Springer,
  \bibinfo{pages}{144--155}.
\newblock


\bibitem[\protect\citeauthoryear{Usman, Lee, Moghe, Zhang, Faloutsos, and
  Kapadia}{Usman et~al\mbox{.}}{2020}]%
        {usman2020social}
\bibfield{author}{\bibinfo{person}{Muhammad Usman}, \bibinfo{person}{Tien-Chi
  Lee}, \bibinfo{person}{Ryhan Moghe}, \bibinfo{person}{Xun Zhang},
  \bibinfo{person}{Petros Faloutsos}, {and} \bibinfo{person}{Mubbasir
  Kapadia}.} \bibinfo{year}{2020}\natexlab{}.
\newblock \showarticletitle{A Social Distancing Index: Evaluating Navigational
  Policies on Human Proximity using Crowd Simulations}.
\newblock In \bibinfo{booktitle}{\emph{Motion, Interaction and Games}}.
  \bibinfo{pages}{1--6}.
\newblock


\bibitem[\protect\citeauthoryear{Usman, Schaumann, Haworth, Berseth, Kapadia,
  and Faloutsos}{Usman et~al\mbox{.}}{2018}]%
        {usman2018interactive}
\bibfield{author}{\bibinfo{person}{Muhammad Usman}, \bibinfo{person}{Davide
  Schaumann}, \bibinfo{person}{Brandon Haworth}, \bibinfo{person}{Glen
  Berseth}, \bibinfo{person}{Mubbasir Kapadia}, {and} \bibinfo{person}{Petros
  Faloutsos}.} \bibinfo{year}{2018}\natexlab{}.
\newblock \showarticletitle{Interactive spatial analytics for human-aware
  building design}. In \bibinfo{booktitle}{\emph{Motion, Interaction, and
  Games}}. \bibinfo{pages}{1--12}.
\newblock


\bibitem[\protect\citeauthoryear{Van Den~Berg, Guy, Lin, and Manocha}{Van
  Den~Berg et~al\mbox{.}}{2011}]%
        {van2011reciprocal}
\bibfield{author}{\bibinfo{person}{Jur Van Den~Berg},
  \bibinfo{person}{Stephen~J Guy}, \bibinfo{person}{Ming Lin}, {and}
  \bibinfo{person}{Dinesh Manocha}.} \bibinfo{year}{2011}\natexlab{}.
\newblock \showarticletitle{Reciprocal n-body collision avoidance}.
\newblock In \bibinfo{booktitle}{\emph{Robotics Research}}.
  \bibinfo{publisher}{Springer}, \bibinfo{pages}{3--19}.
\newblock


\bibitem[\protect\citeauthoryear{Weiss, Litteneker, Jiang, and
  Terzopoulos}{Weiss et~al\mbox{.}}{2019}]%
        {weiss2019position}
\bibfield{author}{\bibinfo{person}{Tomer Weiss}, \bibinfo{person}{Alan
  Litteneker}, \bibinfo{person}{Chenfanfu Jiang}, {and}
  \bibinfo{person}{Demetri Terzopoulos}.} \bibinfo{year}{2019}\natexlab{}.
\newblock \showarticletitle{Position-based real-time simulation of large
  crowds}.
\newblock \bibinfo{journal}{\emph{Computers \& Graphics}}  \bibinfo{volume}{78}
  (\bibinfo{year}{2019}), \bibinfo{pages}{12--22}.
\newblock


\bibitem[\protect\citeauthoryear{WHOSocialDistancing}{WHOSocialDistancing}{[n.d.]}]%
        {WHOSocialDistancing}
WHOSocialDistancing \bibinfo{year}{[n.d.]}\natexlab{}.
\newblock \bibinfo{title}{{WHO: Advice for public}}.
\newblock
  \bibinfo{howpublished}{\url{https://www.who.int/westernpacific/emergencies/covid-19/information/physical-distancing}}.
\newblock
\newblock
\shownote{Accessed: 2020-12-02.}


\bibitem[\protect\citeauthoryear{Yu, Yeung, Tang, Terzopoulos, Chan, and
  Osher}{Yu et~al\mbox{.}}{2011}]%
        {yu2011make}
\bibfield{author}{\bibinfo{person}{Lap~Fai Yu}, \bibinfo{person}{Sai~Kit
  Yeung}, \bibinfo{person}{Chi~Keung Tang}, \bibinfo{person}{Demetri
  Terzopoulos}, \bibinfo{person}{Tony~F Chan}, {and} \bibinfo{person}{Stanley~J
  Osher}.} \bibinfo{year}{2011}\natexlab{}.
\newblock \showarticletitle{Make it home: automatic optimization of furniture
  arrangement}.
\newblock \bibinfo{journal}{\emph{Transactions on Graphics}}
  \bibinfo{volume}{30}, \bibinfo{number}{4} (\bibinfo{year}{2011}).
\newblock


\bibitem[\protect\citeauthoryear{Zhang, Litvinova, Liang, Wang, Wang, Zhao, Wu,
  Merler, Viboud, Vespignani, et~al\mbox{.}}{Zhang et~al\mbox{.}}{2020}]%
        {zhang2020changes}
\bibfield{author}{\bibinfo{person}{Juanjuan Zhang}, \bibinfo{person}{Maria
  Litvinova}, \bibinfo{person}{Yuxia Liang}, \bibinfo{person}{Yan Wang},
  \bibinfo{person}{Wei Wang}, \bibinfo{person}{Shanlu Zhao},
  \bibinfo{person}{Qianhui Wu}, \bibinfo{person}{Stefano Merler},
  \bibinfo{person}{C{\'e}cile Viboud}, \bibinfo{person}{Alessandro Vespignani},
  {et~al\mbox{.}}} \bibinfo{year}{2020}\natexlab{}.
\newblock \showarticletitle{Changes in contact patterns shape the dynamics of
  the COVID-19 outbreak in China}.
\newblock \bibinfo{journal}{\emph{Science}} (\bibinfo{year}{2020}).
\newblock


\bibitem[\protect\citeauthoryear{Zhao and Zeng}{Zhao and Zeng}{2006}]%
        {zhao2006simulated}
\bibfield{author}{\bibinfo{person}{Fang Zhao} {and} \bibinfo{person}{Xiaogang
  Zeng}.} \bibinfo{year}{2006}\natexlab{}.
\newblock \showarticletitle{Simulated annealing--genetic algorithm for transit
  network optimization}.
\newblock \bibinfo{journal}{\emph{Computing in Civil Engineering}}
  \bibinfo{volume}{20}, \bibinfo{number}{1} (\bibinfo{year}{2006}),
  \bibinfo{pages}{57--68}.
\newblock


\end{thebibliography}


\end{document}